\documentclass[onecolumn,showpacs]{revtex4}

\usepackage{graphicx}
\usepackage{dcolumn}
\usepackage{bm}

\input epsf

\begin{document}

\title{\Large Generalized second law of thermodynamics in presence of interacting DBI essence and other dark energies}

\author{\bf  Surajit
Chattopadhyay$^1$$^,$$^2$\footnote{surajit$_{_{-}}2008$@yahoo.co.in,
surajit.chattopadhyay@pcmt-india.net} and Ujjal
Debnath$^2$\footnote{ujjaldebnath@yahoo.com ,
ujjal@iucaa.ernet.in}}

\affiliation{$^1$Department of Computer Application, Pailan
College of Management and Technology, Bengal Pailan Park,
Kolkata-700 104, India.\\
$^2$Department of Mathematics, Bengal Engineering and Science
University, Shibpur, Howrah-711 103, India. }

\date{\today}

\begin{abstract}
In the present work we investigated the validity of the
generalized second law (GSL) of thermodynamics in presence of
interaction between DBI-essence and other three candidates of dark
energy namely modified Chaplygin gas, hessence, tachyonic field
and new agegraphic dark energy. It has been observed that the GSL
breaks down in presence of the interactions. However, the event
horizon remains to be an increasing function of time.
\end{abstract}

\pacs{}

\maketitle

\section{\normalsize\bf{Introduction}}

One of the most alluring observational discoveries of the past
decade has been that the expansion of the universe is speeding up
rather than slowing down [1]. An accelerating universe is strongly
suggested by observations of type Ia high redshift supernovae
provided these behave as standard candles. The case for an
accelerating universe is further strengthened by the discovery of
Cosmic Microwave Background (CMB)[2] and galaxy power spectrum for
large scale structures [3]. These studies indicate that in
spatially flat isotropic universe, about two-thirds of the
critical energy density seems to be stored in a so called dark
energy (DE) component with enough negative pressure responsible
for the currently cosmic accelerating expansion [4]. Strength of
this acceleration depends on the theoretical model employed while
interpreting the data. A wide range of scenarios have been
proposed to explain this acceleration but most of them cannot
explain all the features of the universe or they have so many
parameters that they are difficult to fit. The models which have
been discussed widely in literature are those which consider
vacuum energy (cosmological constant) as DE, introduce a fifth
element and dub it quintessence or scenarios named phantom with
$\omega < -1$, where $\omega=p/\rho$ is a parameter of state
[5].\\\\
An approach to the problem of DE arises from the holographic
principle that states that the number of degrees of freedom
related directly to entropy scales with the enclosing area of the
system [5]. Based on the holographic principle proposed by
Fischler and Susskind [6] several others have studied holographic
model for dark energy (HDE)  [7].  The holographic energy density
is given by $\rho_{\Lambda}=3c^{2}M_{p}^{2}L^{-2}$ in which $c$ is
a free dimensionless parameter and the coefficient 3 is for
convenience. As an application of the holographic principle in
cosmology, it was shown that the consequence of excluding those
degrees of freedom of the system which will never be observed by
the effective field theory gives rise to IR cut-off $L$ at the
future event horizon. Thus, in a universe dominated by DE, the
future event horizon will tend to a constant of the order
$H_{0}^{-1}$, the present Hubble radius [5].\\\\
 In 1973, Bekenstein
[8] assumed that there is a relation between the event of horizon
and the thermodynamics of a black hole, so that the event of
horizon of the black hole is a measure of the entropy of it. This
idea has been generalized to horizons of cosmological models, so
that each horizon corresponds to an entropy. Thus the second law
of thermodynamics was modified in the way that in generalized
form, the sum of all time derivative of entropies related to
horizons plus time derivative of normal entropy must be positive,
i.e. the sum of entropies must be increasing function of time.
Davies [9] investigated the validity of generalized second law
(GSL) for the cosmological models which departs slightly from de
Sitter space. However, it is only natural to associate an entropy
to the horizon area as it measures our lack of knowledge about
what is going on beyond it. Setare [10] investigated the validity
of the generalized second law of thermodynamics for the quintom
model of dark energy. Setare [7] considered the interacting
holographic model of dark energy to investigate the validity of
the generalized second law of thermodynamics in a non-flat
(closed) universe enclosed by the event horizon measured from the
sphere of the horizon $L$. Setare and Shafei [5] showed that for
the apparent horizon the first law is roughly respected for
different epochs while the second law of thermodynamics is
respected, while for $L$ as the system's IR cut-off the first law
is broken and the second law is respected for the special range of
the deceleration parameter.
\\\\
There have been many works aimed at connecting the string theory
with inflation. Various ideas in string theory based on the
concept of branes have proved themselves fruitful. Scenarios where
the inflation is interpreted as the distance between two branes
moving in the extra dimension along a warped throat have given
rise to many interesting studies [11]. One area which has been
well explored in recent years, is inflation driven by the open
string sector through dynamical Dp-branes. This is the so-called
DBI (Dirac-Born-Infield) inflation [12, 13, 14], which lies in a
special class of K-inflation models.  Martin and Yamaguchi [13]
introduced a scalar field model where the kinetic term has a DBI
form and considered that the dark energy scalar field is a DBI
scalar field, for which the action of the field can be written as

\begin{equation}
S_{DBI}=-\int d^{4}xa^{3}
(t)\left[T(\phi)\sqrt{1-\frac{\dot{\phi}^{2}}{T(\phi)}}+V(\phi)-T(\phi)\right]
\end{equation}

where $T(\phi)$ is the tension and $V(\phi)$ is the potential. To
obtain a suitable evolution of the Universe an interaction is
often assumed such that the decay rate should be proportional to
the present value of the Hubble parameter for good fit to the
expansion history of the Universe as determined by the Supernovae
and CMB data [15]. These kind of models describe an energy flow
between the components so that no components are conserved
separately. There are several work on the interaction between dark
energy (tachyon or phantom) and dark matter [16], where
phenomenologically introduced different forms of interaction term.
\\\\
The present paper is an extension of the paper of the present
authors Chattopadhyay and Debnath (2010) [Ref 17]. In the said
paper [17], the interactions between different candidates of dark
energy and DBI-essence model were considered. For the sake of
simplicity in the presentation, the interactions discussed in
reference [17] have been described briefly and subsequently the
GSL has been considered.Organizations of the present paper is as
follows: in section II we have given a brief introduction to the
generalized second law (GSL) of thermodynamics with respect to
cosmology; in section III we have considered the GSL in the
interacting DBI-essence and modified Chaplygin gas (MCG); in
section IV GSL has been viewed in presence of interacting
DBI-essence and hessence; in section V we have investigated GSL in
presence of interacting DBI-essence and tachyonic field.
\\\\

\section{\normalsize\bf{Second law of thermodynamics}}

In the present work, we consider the universe as a flat FRW
universe and take into account that the accelerating universe has
a future event horizon $R_{h}$, which is also named as
cosmological horizon [11]. The radius of observer's event horizon
is given by [5]

\begin{equation}
R_{h}=a \int_{t}^{\infty}\frac{dt}{a}= a
\int_{a}^{\infty}\frac{da}{Ha^{2}}
\end{equation}

To study the generalized second law (GSL) through the universe
under the interaction between tachyonic field and scalar (phantom)
field we would examine the nature of the derivative of the normal
entropy $S$ in presence of interaction. It is a proven fact that
for phantom dominated universe $\dot{S}>0$ and for a quintessence
dominated universe $\dot{S}<0$ [10]. Our target is to answer the
question : \emph{Is $\dot{S}>0$ under the interaction between
DBI-essence and other dark energies}?. The expression for normal
entropy using the first law of thermodynamics is

\begin{equation}
TdS=dE+PdV=(P+\rho)dV+Vd\rho
\end{equation}

Also we know that

\begin{equation}
H^{2}=\frac{1}{3}\rho
\end{equation}

and

\begin{equation}
\dot{H}=-\frac{1}{2}(P+\rho)
\end{equation}

Using $V=\frac{4}{3}\pi R_{h}^{3}$ in equation (3) we get

\begin{equation}\label{1}
TdS=-2\dot{H}dV+Vd\rho=-8\pi R_{h}^{2}\dot{H}dR_{h}+8\pi
R_{h}^{3}dH
\end{equation}

From equation (\ref{1}), it can be obtained that

\begin{equation}
\dot{S}=\frac{8 \pi \dot{H}R_{h}^{2}}{T}
\end{equation}

If the horizon entropy is taken to be $S_{h}=\pi R_{h}^{2}$ , we
get

\begin{equation}
\dot{S}+\dot{S}_{h}=\frac{8 \pi \dot{H}R_{h}^{2}}{T}+2 \pi
R_{h}\dot{R}_{h}\geq 0
\end{equation}

Taking the temperature $T=\frac{1}{2\pi R_{h}}$ and using
$\dot{R_{h}}=HR_{h}-1$  we can write

\begin{equation}
\dot{S}+\dot{S}_{h}=\dot{S}_{X}=16 \pi^{2}\dot{H}R_{h}^{3}+2 \pi
R_{h} (HR_{h}-1)\geq 0
\end{equation}

In the subsequent sections, we would investigate the validity of
the equation (9) in various interacting situations.
\\\\

\section{\normalsize\bf{GSL in presence of interacting DBI-essence and MCG }}

We consider a spatially flat isotropic and homogeneous universe in
the FRW model whose metric is given by

\begin{equation}
ds^{2}=dt^{2}-a^{2}(t)[dr^{2}+r^{2}(d\theta^{2}+sin^{2}d\phi^{2})]
\end{equation}
where, $a(t)$ is the scale factor. The Einstein field equations
are given by (choosing $8\pi G=c=1$)

The energy conservation equation is given by

\begin{equation}
\dot{\rho}+3\frac{\dot{a}}{a} (\rho+p)=0
\end{equation}

If we consider a model consisting of two component mixture, the an
interaction term needs to be introduced. In a two-component model,
we replace $\rho$ and $p$ of equation (11) by $\rho_{total}$ and
$p_{total}$ where

\begin{equation}
\rho_{total}=\rho_{D}+\rho_{X}
\end{equation}

\begin{equation}
p_{total}=p_{D}+p_{X}
\end{equation}

where $\rho_{D}$ and $p_{D}$ denote the density and pressure for
the DBI-essence. The terms $\rho_{X}$ and $p_{X}$ denote the
density and pressure corresponding to the other dark energies.
Therefore, we get

\begin{equation}
3\frac{\dot{a}^{2}}{a^{2}}=(\rho_{D}+\rho_{X})
\end{equation}

\begin{equation}
6\frac{\ddot{a}}{a}=-\left[(\rho_{D}+\rho_{X})+3(p_{D}+p_{X})\right]
\end{equation}

\begin{equation}
(\dot{\rho}_{D}+\dot{\rho_{X}})+3\frac{\dot{a}}{a}[(\rho_{D}+\rho_{X})+(p_{D}+p_{X})]=0
\end{equation}

Assuming gravity to obey four-dimensional general relativity with
a standard Einstein-Hilbert Lagrangian, the density and pressure
for DBI-essence are read as [15]

\begin{equation}
\rho_{D}=(\gamma-1)T(\phi_{D})+V_{D}(\phi_{D})
\end{equation}

\begin{equation}
p_{D}=\left(\frac{\gamma-1}{\gamma}\right)T(\phi_{D})-V_{D}(\phi_{D})
\end{equation}

where, $\phi_{D}$ denotes the field for DBI-essence and the
quantity $\gamma$ is reminiscent of the usual Lorentz factor given
by

\begin{equation}
\gamma=\frac{1}{\sqrt{1-\frac{\dot{\phi}_{D}^{2}}{T(\phi_{D})}}}
\end{equation}

Since we are considering two-component model, we consider the
interaction term $3H\delta\rho_{X}$ and we can write the
conservation equations as

\begin{equation}
\dot{\rho_{D}}+3H(\rho_{D}+p_{D})=3H\delta\rho_{X}
\end{equation}

\begin{equation}
\dot{\rho_{X}}+3H(\rho_{X}+p_{X})=-3H\delta\rho_{X}
\end{equation}

where, $H=\frac{\dot{a}}{a}$ is the Hubble parameter, $\delta$ is
the interaction parameter and rest of the symbols are as explained
earlier. The pressure and density of MCG are given by

\begin{equation}
p_{ch}=A\rho_{ch}-\frac{B}{\rho_{ch}^\alpha}
\end{equation}

and

\begin{equation}
\rho_{ch}=\left[\frac{B}{1+A}+\frac{C}{a^{3(1+\alpha)(1+A)}}\right]^{\frac{1}{1+\alpha}}
\end{equation}

In the reference [17] it was obtained under interaction that

\begin{equation}
\rho_{ch}=\left(\frac{B}{1+A+\delta}+\frac{C}{t^{3m(1+\alpha)(1+A)}}\right)^{\frac{1}{1+\alpha}}
\end{equation}

\begin{equation}
\dot{\phi}_{D}^{2}=2\sqrt{\frac{n-1}{n}}
\left[\frac{m}{t^{2}}+\left(\frac{B}{1+A+\delta}+\frac{C}{t^{3m(1+\alpha)(1+A)}}\right)^{-\frac{\alpha}{1+\alpha}}
\left(B-(1+A)\left(\frac{B}{1+A+\delta}+\frac{C}{t^{3m(1+\alpha)(1+A)}}\right)\right)\right]
\end{equation}

Using expressions (24) and (25), we calculated $\rho_{total}$ and
$p_{total}$ and subsequently calculated $H$ and $\dot{H}$ using
equations (4) and (5). Afterwards, $H$ is used in equation (2) for
event horizon $R_{h}$ which is finally used to generate
$\dot{S}_{X}$ of equation (9). In figure (1) we have presented the
variation of $\dot{S}_{X}$ with the redshift $z=1-t^{-m}$ in the
interacting situation, i.e. with $\delta\neq0$. It is apparent
from figure (1), that $\dot{S}_{X}$ is always remaining in the
negative level. This indicates a violation of the GSL as indicated
in equation (9). In figure (2), we have plotted the change of
$\dot{S}_{X}$ against $z$ in the situation of non-interaction i.e.
$\delta=0$. Here also, it is observed that the GSL is not valid.
 \\
\begin{figure}
\includegraphics[height=1.8in]{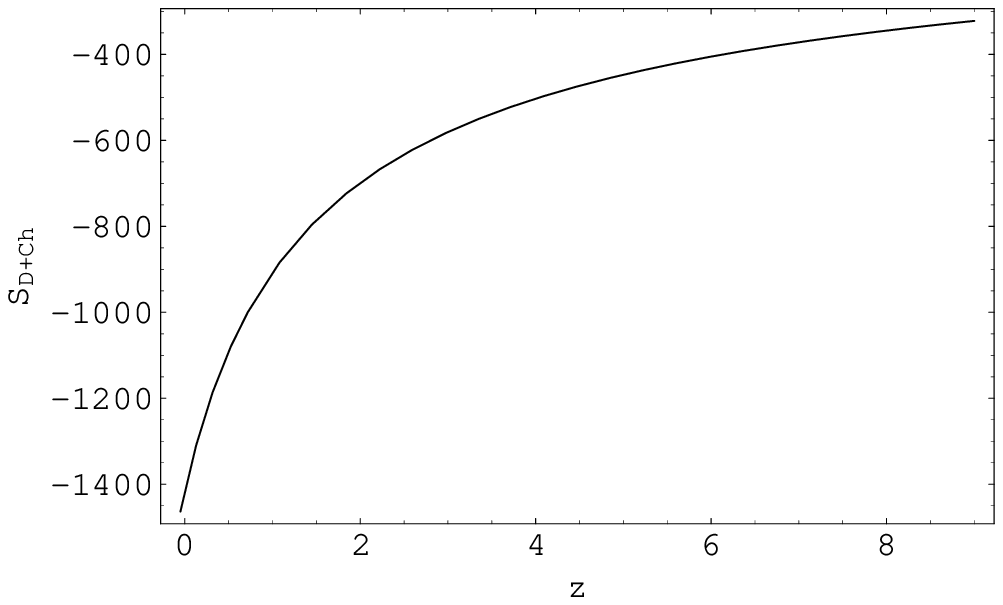}~~~
\includegraphics[height=1.8in]{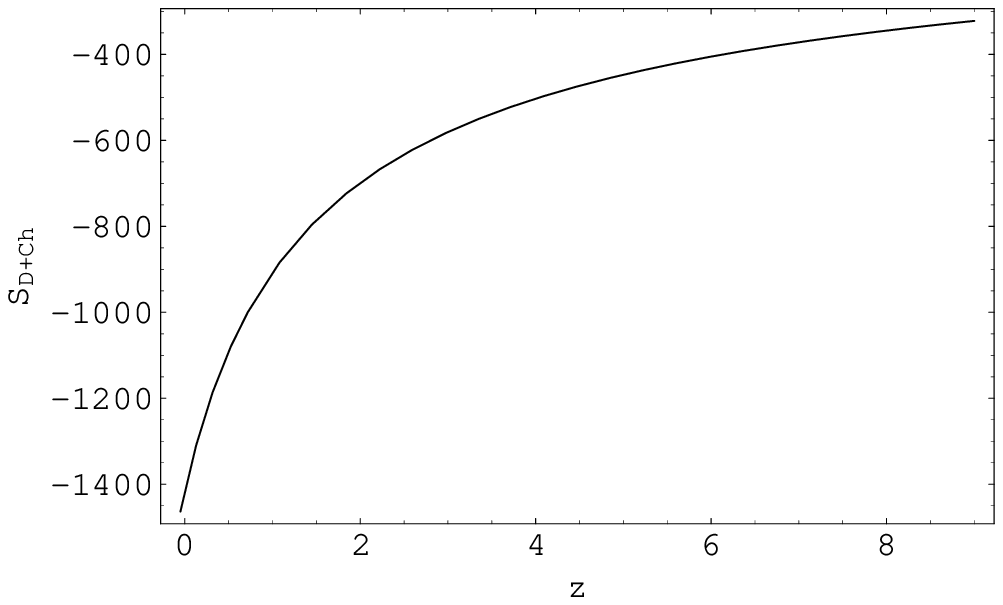}~\\
\vspace{1mm} ~~~~~~~~~~~~Fig.1~~~~~~~~~~~~~~~~~~~~~~~~~~~~~~~~~~~~~~~~~~~~~~~~~~~~~~~~Fig.2\\
\vspace{6mm} Figs. 1 and 2 show the variation of $\dot{S}_{X}$
against $z$ in presence of interaction between DBI-essence and MCG
($\delta=0.05$) and in the case of a mixture of the two dark
energies without interaction ($\delta=0$) .

\vspace{6mm} \vspace{6mm}

\end{figure}

\section{\normalsize\bf{GSL in presence of interacting DBI-essence and hessence}}

\begin{figure}
\includegraphics[height=1.8in]{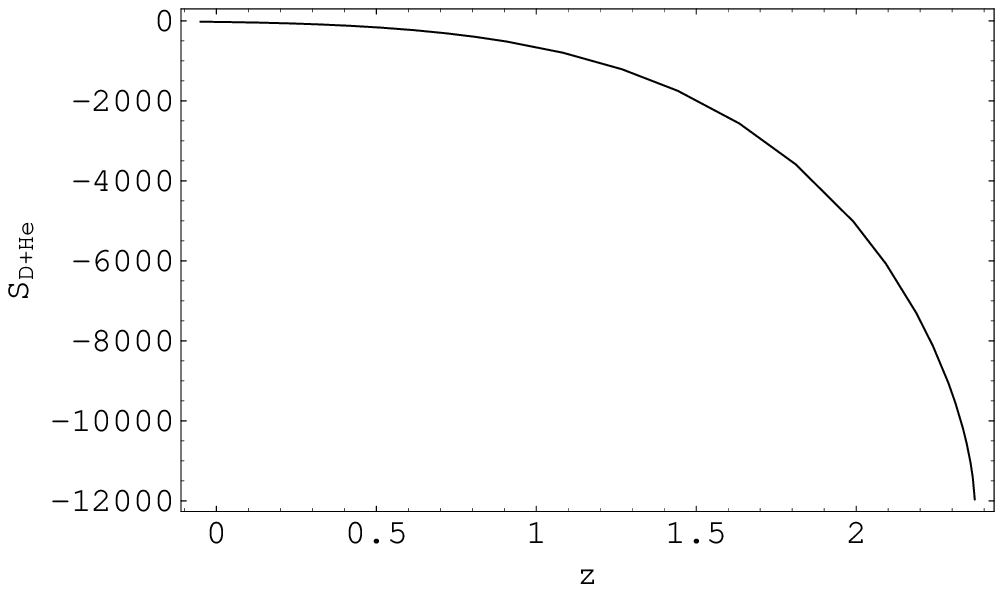}~~~
\includegraphics[height=1.8in]{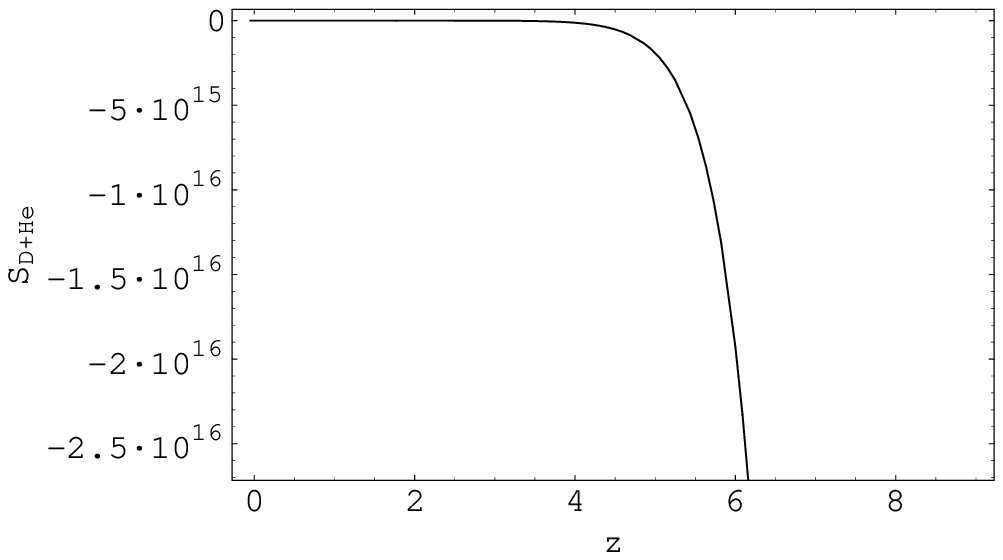}~\\
\vspace{1mm} ~~~~~~~~~~~~Fig.3~~~~~~~~~~~~~~~~~~~~~~~~~~~~~~~~~~~~~~~~~~~~~~~~~~~~~~~~Fig.4\\
\vspace{6mm} Figs. 3 and 4 show the variation of $\dot{S}$ against
$z$ in presence of interaction between DBI-essence and hessence
($\delta=0.05$) and in the case of a mixture of the two dark
energies without interaction ($\delta=0$) .

\vspace{6mm} \vspace{6mm}

\end{figure}

Energy density and pressure in the case of hessence are given by
[17]

\begin{equation}
\rho_{H}=\frac{1}{2}(\dot{\phi}_{H}^{2}-\phi_{H}^{2}\dot{\theta}^{2})+V_{H}(\phi_{H})
\end{equation}

\begin{equation}
p_{H}=\frac{1}{2}(\dot{\phi}_{H}^{2}-\phi_{H}^{2}\dot{\theta}^{2})-V_{H}(\phi_{H})
\end{equation}

where, $\dot{\theta}=\frac{Q}{a^{3}\phi_{H}^{2}}$.\\
Now, taking $T(\phi_{D})=m \phi_{D}^{2}$ and scale factor
$a(t)=t^{n}$. From the field equations it can be found that

\begin{equation}
\frac{n}{t^{2}}=\frac{1}{2}\left(\dot{\phi}_{H}^{2}-\frac{Q^{2}}{t^{6n}\phi_{H}^{2}}+m
\dot{\phi}_{D}^{2} \right)
\end{equation}

Under interaction, the forms of the potentials are found as [17]

\begin{equation}
V_{H}=\frac{1}{4}t^{-3n\delta}\left[(3n-1)Qt^{3n(-2+\delta)}\left(\frac{4\delta}{2-\delta}+
\frac{t^{-2+6n}(2-3n(2+\delta))}{3n\delta-2}\right)+C_{1}\right]
\end{equation}

and

\begin{equation}
V_{D}=\frac{2n^{2}\left(2+3n\delta+\sqrt{\frac{n}{n-1}}(1-3n\delta)\right)}{mt^{2}(2+3n\delta)}+C_{2}t^{3n\delta}
\end{equation}

Like the earlier case, here we calculated the terms involved with
$\dot{S}_{X}$ using the above expressions and plotted its
evolution with the redshift $z$ in both interacting (figure 3) and
non interacting (figure 4) cases. In both of the cases it is
observed that $\dot{S}_{X}$ remains negative and hence GSL is not
valid.
\\

\section{\normalsize\bf{GSL in presence of interacting DBI-essence and tachyonic field}}

The energy density $\rho_{T}$ and pressure $p_{T}$ for tachyonic
field are given by

\begin{equation}
\rho_{T}=\frac{V_{T}(\phi_{T})}{\sqrt{1-\dot{\phi}_{T}^{2}}}
\end{equation}

\begin{equation}
p_{T}=-V_{T}(\phi_{T})\sqrt{1-\dot{\phi}_{T}^{2}}
\end{equation}

Under the interaction, the potentials for tachyonic field and
DBI-essence are obtained as [17]

\begin{equation}
V_{T}=e^{\frac{3nt^{-m}}{m}}t^{-m+3n(1+\delta)}
\end{equation}

\begin{figure}
\includegraphics[height=1.8in]{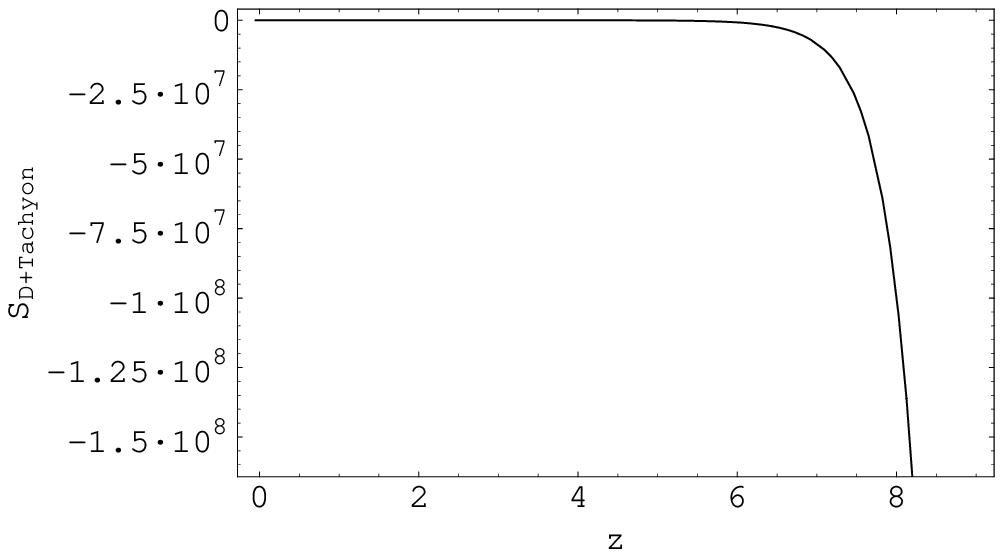}~~~
\includegraphics[height=1.8in]{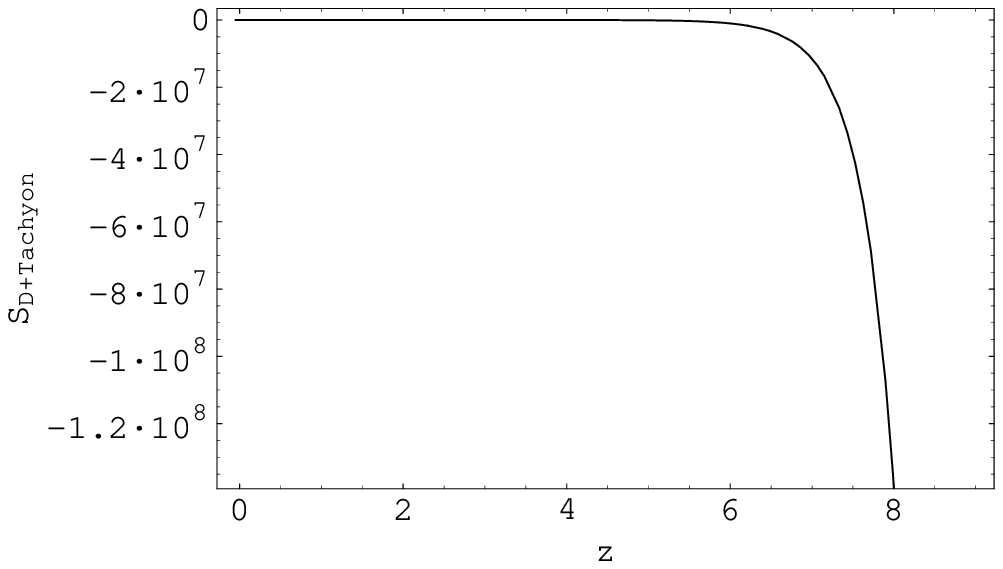}~\\
\vspace{1mm} ~~~~~~~~~~~~Fig.5~~~~~~~~~~~~~~~~~~~~~~~~~~~~~~~~~~~~~~~~~~~~~~~~~~~~~~~~Fig.6\\
\vspace{6mm} Figs. 5 and 6 show the variation of $\dot{S}_{X}$
against $z$ in presence of interaction between DBI-essence and
tachyonic field ($\delta=0.05$) and in the case of a mixture of
the two dark energies without interaction ($\delta=0$) .

\vspace{6mm} \vspace{6mm}

\end{figure}

\begin{equation}
V_{D}=\frac{1}{2}\left(\frac{6n}{t^{2}}+4k_{1}t^{k_{2}}+(3-4k_{1})\sqrt{\frac{k_{1}}{k_{1}-1}}t^{k_{2}}+
\frac{e^{\frac{3nt^{-m}}{m}}t^{-m+3n(1+\delta)}(3+t^{m})}{\sqrt{t^{m}}}\right)
\end{equation}

Like the earlier case, here we calculated the terms involved with
$\dot{S}_{X}$ using the above expressions and plotted its
evolution with the redshift $z$ in both interacting (figure 5) and
non interacting (figure 6) cases. In both of the cases it is
observed that $\dot{S}_{X}$ remains negative and hence GSL is not
valid.
\\

\section{\normalsize\bf{GSL in presence of interacting DBI-essence and new agegraphic dark energy}}

The so-called agegraphic dark energy model was proposed in Cai
(2007) [18], where the energy density is given by

\begin{equation}
\rho_{q}=\frac{3n^{2}m_{p}^{2}}{T^{2}}
\end{equation}

where,

\begin{equation}
T=\int\frac{da}{Ha}
\end{equation}

If we consider a flat FRW universe containing the agegraphic dark
energy and pressureless-matter, the the corresponding Friedman
equation becomes

\begin{equation}
H^{2}=\frac{1}{3m_{p}^{2}}(\rho_{m}+\rho_{q})
\end{equation}

Introducing the fractional energy densities
$\Omega_{i}=\rho_{i}/3m_{p}^{2}H^{2}$ we can get

\begin{equation}
\Omega_{q}=\frac{n^{2}}{H^{2}T^{2}}
\end{equation}

and the equation-of-state parameter $\omega_{q}=p_{q}/\rho_{q}$ is
given by

\begin{equation}
\omega_{q}=-1+\frac{2}{3n}\sqrt{\Omega_{q}}
\end{equation}

A new agegraphic dark energy model was proposed in the reference
[19], where the energy density $\rho_{A}$ is given as

\begin{equation}
\rho_{A}=\frac{3n^{2}m_{p}^{2}}{\eta^{2}}
\end{equation}

where

\begin{equation}
\eta=\int\frac{dt}{a}
\end{equation}

Thus, $\dot{\eta}=1/a$. The corresponding fractional energy
density is given by

\begin{equation}
\Omega_{A}=\frac{n^{2}}{H^{2}\eta^{2}}
\end{equation}

In the present work, we would consider an interaction between
DBI-essence and new agegraphic dark energy. To do so, we consider
equations (14) and (15) and in the present case $\rho_{X}\equiv
\rho_{A}$ and $p_{X}\equiv p_{A}$. Considering the interaction the
equation-of-state parameter and the potential are obtained as [17]

\begin{equation}
\omega_{A}=-1-\delta+\frac{2}{3n}\frac{\sqrt{\Omega_{A}}}{a}
\end{equation}

\begin{equation}
V_{D}=\frac{1}{2}\left(\frac{6k_{1}}{t^{2}}+(1-t^{k_{2}})\left(4k_{3}-\sqrt{\frac{k_{3}}{k_{3}-1}}(4k_{3}-3)\right)
-\frac{3}{k_{1}}(k_{1}-1)^{2}(-2+3k_{1}(2+\delta))n^{2}m_{p}^{2}t^{2(k_{1}-1)}\right)
\end{equation}

\begin{figure}
\includegraphics[height=2.0in]{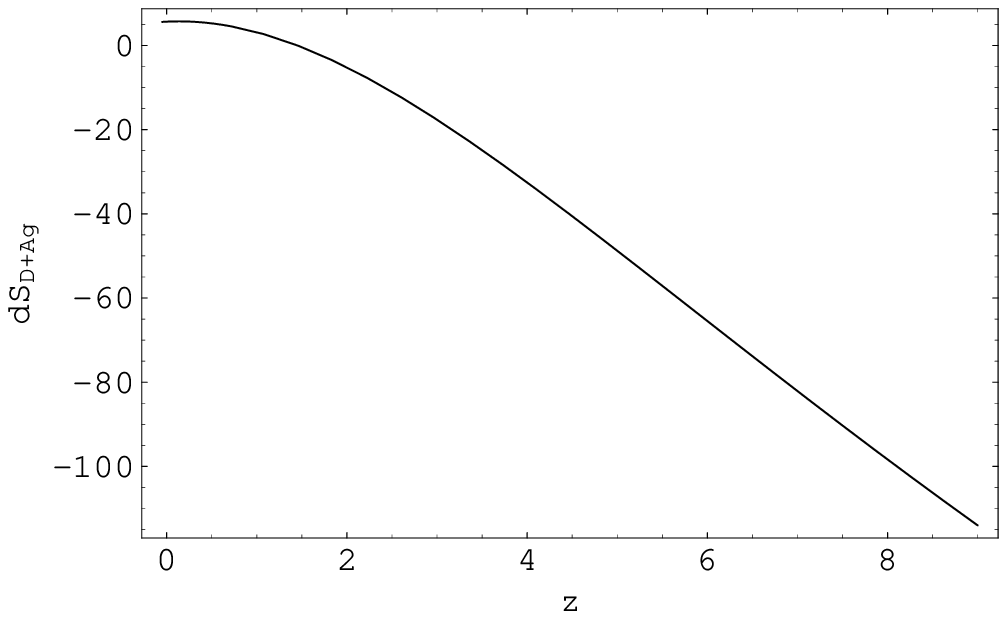}~~~~~~
\includegraphics[height=2.0in]{Fig7.eps}~\\
\vspace{1mm} ~~~~~~~~~~~~Fig.7~~~~~~~~~~~~~~~~~~~~~~~~~~~~~~~~~~~~~~~~~~~~~~~~~~~~~~~~Fig.8\\
\vspace{6mm} Figs. 7 and 8 show the variation of $\dot{S}_{X}$
against $z$ in presence of interaction between DBI-essence and new
agegraphic dark energy($\delta=0.05$) and in the case of a mixture
of the two dark energies without interaction ($\delta=0$) .

\vspace{6mm}

\end{figure}

The $\dot{S}_{X}$ using the above expressions and plotted its
evolution with the redshift $z$ in both interacting (figure 7) and
non interacting (figure 7) cases. In both of the cases it is
observed that $\dot{S}_{X}$ remains negative and hence GSL is not
valid.
\\

\section{\normalsize\bf{Concluding remarks}}
In the present paper we have considered total entropy as the
entropy of a cosmological event horizon plus the entropy of the
quintom fluid. We have investigated the validity of GSL in some
interacting situations. In all of the cases we have observed that
the time derivative of the total entropy is remaining at the
negative level. This means that the total entropy is a decreasing
function of time in the interacting situations considered in this
paper. This means that the GSL breaks down. In an earlier work,
[20] has shown that the GSL breaks down in the situation of
interaction between holographic dark energy and dark matter. The
present work proves the breaking down of GSL in different other
interactions between candidates of dark energies. However, in all
the cases, we have seen that $\dot{R}_{h}$ is always positive.
This means that time derivative of future event horizon is
positive and it indicates that future event horizon is increasing
with time. However, total entropy is decreasing under interaction.
It should be further noted that even in the case of
non-interaction between two candidates of dark energies the GSL is
breaking down.
\\

{\bf References:}\\

$[1]$  A.G.Riess et al., {\it Astron.J.} {\bf 116} 1009 (1998);
S.Perlmutter et al., {\it Astrophys.J.} {\bf 517} 565 (1999).\\
$[2]$ D.N.Spergel et al., {\it Astrophys.J.Suppl.} {\bf 148} 175 (2003).\\
$[3]$ M.Tegmark et al., {\it Phys.Rev.D} {\bf 69} 103501 (2004).\\
$[4]$ A.G.Riess et al., {\it Astron.J.} {\bf 116} 1009 (1998). \\
$[5]$ M. R. Setare and S. Shafei, {\it JCAP} {\bf 09},
stacks.iop.org/JCAP/2006/i=09/a=011 (2006).\\
$[6]$  L. Susskind,  {\it J. Math. Phys.} {\bf 36} 6377 (1995)\\
$[7]$ S. Chattopadhyay and U. Debnath, {\it Astrophys. Space.
Sci.} {\bf 319} 183 (2009); M. R. Setare, {\it JCAP} {\bf 23}
stacks.iop.org/JCAP/2007/i=01/a=023 (2007); M. R. Setare, {\it
Phys. Lett. B} {\bf 648} 329 (2007).\\
$[8]$ J.D. Bekenstein, {\it Phys. Rev. D} {\bf 7} 2333(1973).\\
$[9]$ H. Mohseni Sajadi, gr-qc/0512140.\\
$[10]$ M.R. Setare, {\it Phys. Lett. b}, {\bf 641} 130 (2006).\\
$[13]$ J. Martin and M. Yamaguchi, {\it Phys. Rev. D} {\bf 77}
123508 (2008).\\
$[11]$ S. Kachru et al., {\it JCAP} {\bf 10} 013 (2003); E.
Silverstein and D. Tong, {\it Phys. Rev. D}
{\bf 70} 103505 (2004).\\
$[12]$ B. Gumjudpai and J. Ward, [arXiv:0904.0472v1].\\
$[14]$ J. E. Lidsey and I. Huston, {\it JCAP} {\bf 0707} 002
(2007); D. Baumann and L. McAllister, {\it Phys. Rev. D} {\bf 75}
123508 (2007).\\
$[15]$ M. S. Berger, H. Shojaei, {\it Phys. Rev. D} {\bf 74}
043530
(2006).\\
$[16]$ R. Herrera, D. Pavon and W. Zimdahl, {\it Gen. Rel. Grav.}
{\bf 36} 2161 (2004); R. G. Cai and A. Wang, {\it JCAP} {\bf 0503}
002 (2005); Z. K. Guo, R. G. Cai and Y. Z. Zhang, {\it JCAP}
{\bf 0505} 002 (2005); T. Gonzalez and I. Quiros, [gr-qc/0707.2089].\\
$[17]$ S. Chattopadhyay and U. Debnath, {\it International J. of
Theoretical Physics} {\bf 49} 1465 (2010).\\
$[18]$ R-G. Cai, {\it Phys.Lett. B} {\bf 657} 228
(2007)[arXiv:0707.4049].\\
$[19]$ H. Wei and R-G. Cai, {\it Phys. Lett. B} {\bf 660} 113
(2008)[arXiv:0708.0884].\\
$[20]$ G. Izquierdo and D. Pavon, {\it Phys. Lett. B} {\bf 633}
420 (2006).\\
\end{document}